\documentclass[conference,10pt,a4paper]{IEEEtran}
\IEEEoverridecommandlockouts
\usepackage{amsmath}

\usepackage{amsopn}
\DeclareMathOperator{\diag}{diag}
\DeclareMathOperator{\var}{var}

\usepackage{amssymb}

\usepackage{amsfonts}

\usepackage{graphicx}

\usepackage{caption3} 
\DeclareCaptionOption{parskip}[]{} 
\usepackage[small]{caption}

\usepackage{subcaption}

\usepackage{algorithm}
\usepackage[]{algpseudocode}
\makeatletter
\def\BState{\State\hskip-\ALG@thistlm}
\makeatother
\usepackage{algcompatible}

\newcommand*\diff{\mathop{}\!\mathrm{d}}

\usepackage{cite}

\usepackage{color}
\ifodd 1
\newcommand{\com}[1]{\textbf{\color{blue} (COMMENT: #1)}} 
\else

\newcommand{\com}[1]{}
\fi

\begin{document}
\bibliographystyle{IEEEtran}
\bstctlcite{IEEEexample:BSTcontrol}
	
\title{Placement Optimization and Power Control \\in Intelligent Reflecting Surface Aided \\Multiuser System}

\author{Bifeng~Ling, Jiangbin~Lyu,~\IEEEmembership{Member,~IEEE}, and~Liqun~Fu,~\IEEEmembership{Senior Member,~IEEE}%
\thanks{The authors are with the School of Informatics, Xiamen University, China 361005 (email: bifengling@stu.xmu.edu.cn; \{ljb, liqun\}@xmu.edu.cn). \textit{Corresponding author: Jiangbin Lyu}.}}


\maketitle

\begin{abstract}
Intelligent reflecting surface (IRS) is a new and revolutionary technology capable of reconfiguring the wireless propagation environment by controlling its massive low-cost passive reflecting elements. 
Different from prior works that focus on optimizing IRS reflection coefficients or single-IRS placement, we aim to maximize the minimum throughput of a single-cell multiuser system aided by multiple IRSs, by joint multi-IRS placement and power control at the access point (AP), which is a mixed-integer non-convex problem with drastically increased complexity with the number of IRSs/users.
To tackle this challenge, a ring-based IRS placement scheme is proposed along with a power control policy that equalizes the users' non-outage probability.
An efficient searching algorithm is further proposed to obtain a close-to-optimal solution for arbitrary number of IRSs/rings.
Numerical results validate our analysis and show that our proposed scheme significantly outperforms the benchmark schemes without IRS and/or with other power control policies.
Moreover, it is shown that the IRSs are preferably deployed near AP for coverage range extension, while with more IRSs, they tend to spread out over the cell to cover more and get closer to target users.
%
\end{abstract}


\section{Introduction}
Intelligent reflecting surface (IRS) is a novel technology capable of reconfiguring the wireless signal propagation by controlling its massive low-cost passive reflecting elements, thereby realizing the concept of smart radio environments\cite{QingqingWuTowardsSmart}. 
Compared with the conventional active relaying/beamforming, IRS does not require any active RF chain for signal transmission/reception but simply leverages passive wave reflection, thus having much lower hardware cost and energy consumption yet operating spectral efficiently in full-duplex (FD) without the need of costly self interference cancellation\cite{MDRenzoIRSvsRelay}.
The advantages of IRS have attracted a great deal of research interest in investigating IRS-aided wireless systems (see, e.g., the recent overview/surveys\cite{QingqingWuTowardsSmart,MDRenzoIRSvsRelay,ShiminGongSurvey,QingqingWuTutorial} and the references therein). The majority of existing works aim to optimize the system performance at the link level with one or more IRSs deployed at fixed locations, which show that the IRS-aided system can achieve significant energy efficiency\cite{ChongwenHuangIRSenergyEfficiency} and/or spectral efficiency improvement over the traditional system without IRS\cite{TWCQingqingWuIntelligentReflecting}. 

Besides the active/passive beamforming optimization, another stream of research focuses on the large-scale deployment of IRSs in a hybrid active/passive wireless network\cite{JiangbinLyuHybrid} involving multiple users/access points (APs) aided by multiple IRSs, where a critical issue is the multi-IRS placement optimization which affects the large-scale channel statistics, user association and hence also the system-level performance\cite{QingqingWuTutorial}.
Since IRSs have much lower cost compared with active APs/relays\cite{QingqingWuTowardsSmart}\cite{MDRenzoIRSvsRelay}, they can be much more densely deployed in order to effectively alter the signal propagation in the network, which, however, leads to drastically increased complexity in solving the large-scale multi-IRS placement problem.
Given a total number of reflecting elements, there are various IRS deployment strategies by placing these elements at different locations, e.g., near AP/users or both\cite{you2020deploy}, or dividing them into smaller-size IRSs that are distributed in the network\cite{ShuowenZhangCapacityRegion}\cite{YunlongCaiIntelligentReflecting}.
However, the IRS locations are assumed to be fixed and given in the above works\cite{you2020deploy,ShuowenZhangCapacityRegion,YunlongCaiIntelligentReflecting} and not explicitly optimized.
The authors in \cite{XidongMuJointDeployment} focus on the deployment optimization of one single IRS under different multiple access schemes to maximize the weighted sum rate, whereas multi-IRS placement is yet to be considered.
In our prior work\cite{JiangbinLyuSpatialThroughput}, the spatial throughput of a single-cell multi-user system aided by distributed IRSs located at random locations is characterized, which is compared favorably with the conventional system aided by distributed relays but with significantly reduced active antennas. However, \cite{JiangbinLyuSpatialThroughput} optimizes only the IRS deployment range in the cell instead of detailed multi-IRS placement optimization, and power control is not considered which affects the signal propagation range and thus couples with IRS placement for providing wireless coverage.

In this paper, we focus on the multi-IRS placement optimization in a single-cell multiuser system along with the downlink AP power control, and aim to maximize the minimum throughput of all user equipments (UEs) in order to provide fairness and/or gauge the maximum supported UE density with minimum rate requirements.
Similar max-min fairness is investigated in \cite{XieMaxMinTWC}, whereas IRS placement is not explicitly considered. 
Note that we choose the UE's (average) throughput with a certain non-outage probability (NOP) requirement as the performance metric instead of its instantaneous rate, since we are interested in the system-level throughput optimization/capacity planning in the long run, and aim to obtain a general IRS placement/AP power control solution that pertains to statistical UE distributions/channel state information (CSI).
However, even considering only macro-decisions of IRS deployment and ``slow" power control based on statistical CSI, the resulted problem is still difficult to solve due to the non-convex constraint of NOP requirement and the integer constraint of UE-to-IRS association, with drastically increased complexity as the number of IRSs/UEs increases.


To tackle the above challenge and characterize/optimize the system-level performance, 
we first abstract the link-level details by deriving the overall channel statistics for the AP-UE communication assisted by IRS reflect beamforming, based on which we obtain a closed-form approximation of the required transmit power (TP) to satisfy the NOP requirement under given locations of the target UE and its serving IRS.
In addition, based on the derived channel statistics, we investigate the impact of IRS deploying position on the coverage range extension in the cell, which suggests two desirable modes of IRS deployment, i.e., near-AP deployment with long range coverage, or near-UE deployment with local coverage.
Thereby, we propose a \textit{ring-based IRS placement scheme} where the UEs are grouped into ring regions based on the AP-UE distance and served by either near-AP or near-UE IRSs, as shown in Fig. \ref{IRS}, along with a \textit{power control policy} that equalizes the NOP in a given UE region subject to the total TP constraint.
As a result, the problem complexity is greatly reduced and we are able to obtain the average max-min throughput by searching over the power allocation ratio, partitioning distance, and the number of IRSs deployed for each ring region, when the number of IRSs/rings is small.
An efficient searching algorithm is further proposed which can obtain a close-to-optimal solution for arbitrary number of IRSs/rings.
Numerical results validate our analysis and show that our proposed scheme significantly outperforms the baseline scheme without IRS, as well as the benchmark schemes with other power control policies.
Moreover, it is found that for a small number of IRSs, they are preferentially deployed near AP owing to their wide coverage range, and the cell-edge UEs are covered first to achieve max-min fairness.
As the number of IRSs increases, the IRSs first tend to spread out over the cell to cover more UEs, and then get denser to get closer to their served UEs and achieve higher throughput.


\section{System Model}
\begin{figure}
	\centering
	\includegraphics[width=0.95\linewidth,  trim=0 0 0 0,clip]{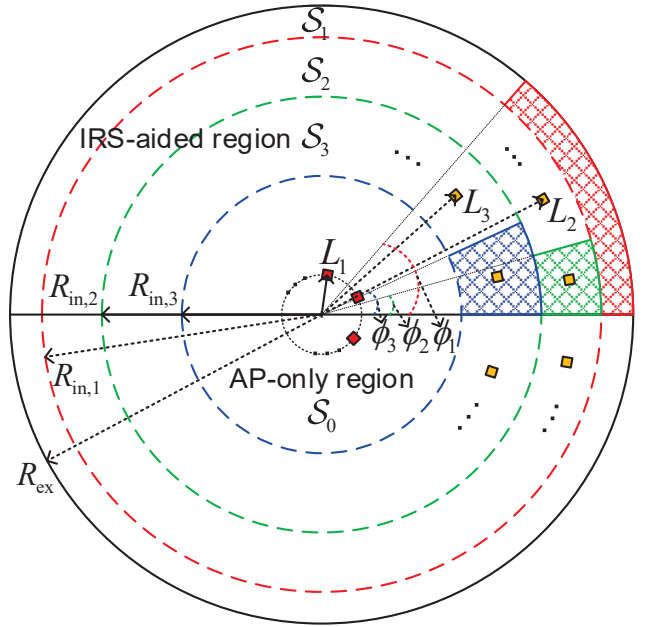}
	\caption{Multi-IRS enhanced wireless coverage in a single cell.\vspace{-2ex}}\label{IRS}
\end{figure}

Consider a single-cell multiuser system with one AP serving a set of $K$ UEs, denoted by $\mathcal{K}\triangleq \{1,\cdots,K\}$, which are uniformly and randomly distributed\footnote{The assumption of uniformly random UE distribution serves as a good baseline to evaluate the general network performance, while our proposed method can be extended to account for the non-uniform UE distribution case by placing more IRSs near the regions with higher UE density.} in the disc cell region of radius $R_\textrm{ex}$ meters (m) centered at AP, as shown in Fig. \ref{IRS}, with an average UE density of $\lambda \triangleq K/( \pi R_{\textrm{ex}}^2 ) $.
We consider the downlink communication from AP to UEs, whereas the results obtained can be similarly applied to the uplink communication as well.
Orthogonal frequency division multiple access (OFDMA) is considered whereby the total bandwidth $B$ is equally partitioned into $n_{\textrm{b}}$ sub-bands and the time frame $T$ is equally partitioned into $n_{\textrm{t}}$ slots. For simplicity, we assume $n_{\textrm{b}} \times n_{\textrm{t}} = K$ and that each UE is randomly allocated with an orthogonal resource block (RB).

A set of $M$ IRSs, denoted by $\mathcal{M}\triangleq \{1,\cdots,M\}$, are deployed to assist the AP-UE communications. Denote the set of IRS horizontal locations as $\mathcal{W} \triangleq \{ \bold{w}_m \in \mathbb{R}^2 | m \in \mathcal{M}, R_{\textrm{min}}\leq\| \bold{w}_m \| \leq R_{\textrm{ex}} \}$, where $\mathbb{\bold{w}}_m$ is the two-dimensional (2-D) coordinate of an IRS $m \in \mathcal{M}$, and $\|\cdot\|$ denotes the Euclidean norm. 
To maximize the passive beamforming gain of the IRS to each served UE, we assume that its served UEs are assigned in orthogonal-time RBs, i.e., time division multiple access (TDMA) or time sharing is adopted for the UEs served by the same IRS.\footnote{It is shown in \cite{ZhengBeiXiongNOMAorOMA} that for IRS-aided multiple access, the TDMA scheme is in general superior over the FDMA scheme due to the hardware limitation of IRS passive reflection, which can be made time-selective, but not frequency-selective \cite{QingqingWuTowardsSmart}. Other multiple access schemes are left for future investigation.}
By denoting $K_m$ as the number of UEs served by IRS $m$ within one time frame, we have $K_m \leq n_{\textrm{t}}$.
For simplicity and ease of implementation, we assume that each UE $k$ can be served by at most one IRS. Let $a_{m, k} = 1$ represent the case where UE $k$ is served by IRS $m$, and $a_{m, k} = 0$ otherwise.

\subsection{Channel Model}\label{ChannelModel}
Assume that AP and UEs are each equipped with a single antenna, while each IRS has $N$ reflecting elements.
The baseband equivalent channel from AP to UE $k$ is denoted by $h_{\textrm{d}, k}\in \mathbb{C}$, where $\mathbb{C}$ denotes the set of complex numbers.
If UE $k$ is also served by IRS $m_k$, the baseband equivalent channels from AP to IRS $m_k$, and from IRS $m_k$ to UE $k$ are denoted by $\bold h_{\textrm{i}, k} \triangleq [h_{\textrm{i},k,1},\cdots,h_{\textrm{i},k,N}]^T\in \mathbb{C}^{N \times 1}$ and $\bold h_{\textrm{r}, k} \triangleq [h_{\textrm{r},k,1},\cdots,h_{\textrm{r},k,N}]^T\in \mathbb{C}^{N \times 1}$ respectively, where $[\cdot]^T$ denotes the matrix transpose. 
Let $\boldsymbol\theta \triangleq [\theta_1, \cdots,\theta_N]$ and furthermore denote $\boldsymbol\Theta \triangleq \diag\{ [e^{j\theta_1}, \cdots,e^{j\theta_N}]\}$ (with $j$ denoting the imaginary unit) as the phase-shifting matrix of IRS $m_k$, where $\theta_n\in[0,2\pi)$ is the phase shift by element $n$ on the incident signal,\footnote{In this paper, we assume (maximum) unit amplitude for each reflection coefficient to maximize the IRS beamforming gain to its served UE \cite{TWCQingqingWuIntelligentReflecting}.} and $\diag\{\bold x\}$
denotes a diagonal matrix with each diagonal element being the corresponding element in $\bold x$. The cascaded AP-IRS-UE channel is then modeled as a concatenation of three components, namely, AP-IRS link, IRS reflecting with phase shifts,
and IRS-UE link, given by \cite{TWCQingqingWuIntelligentReflecting}
\begin{equation}
h_{\textrm{ir}, k}\triangleq \bold h_{\textrm{i}, k}^T \boldsymbol\Theta \bold h_{\textrm{r}, k}.
\end{equation}

Assume that the cascaded channel phase $\angle(h_{\textrm{i},k,n}h_{\textrm{r},k,n})$ via each IRS element $n=1,\cdots,N$ can be obtained via IRS-customized channel estimation methods\cite{ZhengbeiXiongIRSenhancedOFDM}.
The IRS then adjusts the phase shift $\boldsymbol\theta$ such that the $N$ reflected signals are of the same phase at its served UE's receiver by setting $\theta_n=-\angle(h_{\textrm{i},k,n}h_{\textrm{r},k,n}), n=1,\cdots,N$.
As a result, we have
\begin{equation}\label{hirk}
|h_{\textrm{ir}, k}|=|\bold h_{\textrm{i}, k}|^T |\bold h_{\textrm{r}, k}|=\textstyle{\sum}_{n=1}^N |h_{\textrm{i},k,n}| |h_{\textrm{r},k,n}|,
\end{equation}
where $|\bold x|$ takes the element-wise amplitude of vector $\bold x$.
Assume that the AP-UE channel phase $\angle h_{\textrm{d}, k}$ is also known and the IRS can perform a common phase-shift such that $h_{\textrm{ir}, k}$ and $h_{\textrm{d}, k}$ are co-phased and hence coherently combined at the UE\cite{TWCQingqingWuIntelligentReflecting},\footnote{Note that coherent combining can be done for each individual UE alone, without solving a global optimization problem involving all UEs.} with the overall channel amplitude $Z_k \triangleq |h_{\textrm{ir}, k}|+|h_{\textrm{d}, k}|$.

For the AP-IRS, IRS-UE and AP-UE links, we assume a block-fading channel which consists of distance-dependent path-loss with path-loss exponent $n_0 \geq 2$ and an additional random term $\xi$ accounting for small-scale fading.\footnote{Shadowing effect can also be considered by treating it as equivalent random perturbation in the UEs' locations, which is ignored in this work for simplicity.}
The channel power gain of the direct AP-UE $k$ link is thus given by
\begin{equation}\label{gau}
|h_{\textrm{d}, k}|^2 \triangleq g_{\textrm{d}, k}\xi_{\textrm{d}, k}=\alpha_0 (r_k^2+H_\textrm{A}^2)^{-n_0/2}\xi_{\textrm{d}, k},
\end{equation}
where $g_{\textrm{d}, k}$ denotes the average channel power gain, $r_k$ denotes the AP-UE $k$ horizontal distance, $H_\textrm{A}$ denotes the AP height, and $\alpha_0=(\frac{4\pi f_c}{c})^{-2}$ denotes the average channel power gain at a reference distance of 1 m, with $f_c$ denoting the carrier frequency, and $c$ denoting the speed of light; and
$\xi_{\textrm{d}, k} \sim \textrm{Exp}(1)$ is an exponential random variable (RV) with unit mean accounting for the small-scale Rayleigh fading.
Accordingly, $|h_{\textrm{d}, k}|$ follows the Rayleigh distribution $\mathcal{R}( \delta )$ with scale parameter $\delta \triangleq \sqrt{ g_{\textrm{d}, k}/2 }$.
Similarly, the channel power gains from AP to the $n$-th element of the serving IRS $m_k$, and from the latter to UE $k$ are respectively given by
\begin{equation}\label{gai}
|h_{\textrm{i},k,n}|^2 \triangleq g_{\textrm{i}, k}\xi_{\textrm{i},k,n}=\alpha_0 \big(l_{m_k}^2+(H_\textrm{A}-H_\textrm{I})^2\big)^{-n_0/2}\xi_{\textrm{i},k,n},
\end{equation}
\begin{equation}\label{giu}
|h_{\textrm{r},k,n}|^2 \triangleq g_{\textrm{r}, k}\xi_{\textrm{r},k,n}=\alpha_0 \big(d_k^2+H_\textrm{I}^2\big)^{-n_0/2}\xi_{\textrm{r},k,n},
\end{equation}
where $g_{\textrm{i},k}$ and $g_{\textrm{r},k}$ denote the average channel power gains while $l_{m_k}$ and $d_k$ denote the horizontal distances, respectively, and $H_\textrm{I}$ denotes the height of the IRS.\footnote{For the purpose of exposition, we consider far-field propagation for all links, and accordingly assume $H_\textrm{A}\geq 1$ m and $H_\textrm{I}\geq 1$ m, which also avoid unbounded power gain when the horizontal distance $l_{m_k}$ or $d_k$ becomes zero.}
We also assume Rayleigh faded channel for the AP-IRS and IRS-UE links, i.e., $\xi_{\textrm{i},k,n}, \xi_{\textrm{r},k,n}\stackrel{\textrm{dist.}}{=}\xi\sim \textrm{Exp}(1)$.\footnote{The proposed analytical method in this paper can be extended to other fading channel models such as Rician fading.}
Therefore, we have $|h_{\textrm{i},k,n}| \sim \mathcal{R}\big(\sqrt{g_{\textrm{i},k}/2}\big)$ and $|h_{\textrm{r},k,n}| \sim \mathcal{R}\big(\sqrt{g_{\textrm{r},k}/2}\big)$.
Finally, the AP-UE $k$'s channel power gain is given by
\begin{align}
g_k \triangleq \begin{cases}
|h_{\textrm{d}, k}|^2, \quad &\textrm{AP-only}, \\
Z_k^2=(|h_{\textrm{ir}, k}|+|h_{\textrm{d}, k}|)^2, \quad &\textrm{assisted by IRS}.
\end{cases}
\end{align}

\subsection{Channel Statistics in IRS-Aided Communication}
Assume that the fading channels $h_{\textrm{d}, k}$, $h_{\textrm{i},k,n}$ and $h_{\textrm{r},k,n}$, $n=1,\cdots,N$ are independent.
Then for the AP-IRS-UE signal that traverses through element $n$, the channel amplitude is subject to double-Rayleigh fading given by
\begin{equation}
|h_{\textrm{ir},k,n}|\triangleq |h_{\textrm{i},k,n}| |h_{\textrm{r},k,n}|,
\end{equation}
whose mean and variance are respectively given by \cite{JiangbinLyuSpatialThroughput}
\begin{equation}
\mathbb{E}\{|h_{\textrm{ir},k,n}|\}\triangleq\frac{\pi}{4}\sqrt{g_{\textrm{i},k}g_{\textrm{r},k}},
\end{equation}
\begin{equation}
\var\{|h_{\textrm{ir},k,n}|\}\triangleq(1-\pi^2/16)g_{\textrm{i},k}g_{\textrm{r},k}.
\end{equation}

Since the channel amplitudes $|h_{\textrm{ir},k,n}|$, $n=1,\cdots,N$ are independently and identically distributed (i.i.d.),
by the central limit theorem (CLT), the composite amplitude for the AP-IRS-UE channel for practically very large\footnote{We consider electrically small IRSs \cite{MDRenzoIRSvsRelay}, where each reflecting element is typically bounded within a square region of side length around $1/5$ wavelength. Therefore, to fit into the size of 1 m$^2$ at $f_c=2$ GHz, we have $N>1000$, while it can be even larger at higher frequency.} $N$ can be approximated by the Gaussian distribution \cite{JiangbinLyuHybrid}, i.e.,
\begin{small}
\begin{align}
&|h_{\textrm{ir},k}|=\textstyle{\sum}_{n=1}^N |h_{\textrm{ir},k,n}|\stackrel{\textrm{approx.}}{\sim} \mathcal{N}\big(N\mathbb{E}\{|h_{\textrm{ir},k,n}|\},N\var\{|h_{\textrm{ir},k,n}|\}\big)\notag\\
&=\mathcal{N}\big(N\frac{\pi}{4}\sqrt{g_{\textrm{i},k}g_{\textrm{r},k}},\quad N(1-\pi^2/16)g_{\textrm{i},k}g_{\textrm{r},k}\big). \label{haiu}
\end{align}
\end{small}\noindent
 
Finally, the composite channel amplitude $Z_k$ is the sum of a Gaussian RV and an independent Rayleigh RV, hence the mean and variance of $Z_k^2$ are respectively given by
\begin{align}\small
&\mathbb{E}\{Z_k^2\} \triangleq \mathbb{E}\{(|h_{\textrm{ir},k}|+|h_{\textrm{d}, k}|)^2\} \notag \\
&=G_\textrm{bf} g_{\textrm{i},k}g_{\textrm{r},k}+N\frac{\pi}{4}\sqrt{\pi g_{\textrm{i},k}g_{\textrm{r},k}g_{\textrm{d},k}}+g_{\textrm{d},k}, \label{E_Z2}
\end{align}
where $G_\textrm{bf}\triangleq \frac{\pi^2}{16}N^2+\big(1-\frac{\pi^2}{16}\big)N$, and 
\begin{align}
\var\{Z_k^2\} \triangleq \mathbb{E}\{Z_k^4\}-(\mathbb{E}\{Z_k^2\})^2, \label{var_Z2}
\end{align}
which can be obtained from the first four moments of the Gaussian distributed $|h_{\textrm{ir},k}|$ and the Rayleigh distributed $|h_{\textrm{d}, k}|$, whose detailed expression is omitted here for brevity.

\subsection{SNR, Non-Outage Probability, and Throughput}\label{NonOutageProbability}
Denote $p_k$ as the downlink TP to UE $k$ in its allocated RB. The instantaneous signal-to-noise ratio (SNR) at UE $k$'s receiver is given by
\begin{equation}
\gamma_k \triangleq p_k g_k / W, \label{SNRdef}
\end{equation}
where the noise is assumed to be additive white Gaussian noise (AWGN) with power $W \triangleq N_0 b_0$, with $b_0 \triangleq B/n_{\textrm{b}}$ denoting the RB bandwidth, and $N_0$ denoting the noise power density.
As a result, the instantaneous achievable rate within UE $k$'s allocated RB in bits/second/Hz (bps/Hz) is given by
\begin{equation}
R_k \triangleq \log_2( 1+ \gamma_k). \label{InstantaneousThroughputDef}
\end{equation}
Denote $\bar{R}$ as the minimum instantaneous rate in bps/Hz required by an UE in order not to be in \textit{outage}. 
Then the non-outage probability (NOP) is defined as
\begin{align}\label{PnonOutDef}
\textrm{P}_{\textrm{no}, k}& \triangleq \mathbb{P}\{ R_k \geq \bar{R} \} = \mathbb{P}\{ \log_2( 1+ \gamma_k ) \geq \bar{R} \} \notag \\
&= \mathbb{P}\{ \gamma_k \geq 2^{ \bar{R} } - 1 \} = \mathbb{P}\{ \gamma_k \geq \eta_0 \},
\end{align}
where $\eta_0 \triangleq 2^{ \bar{R} }-1$ denotes the corresponding SNR threshold, and UE $k$'s throughput is given by
\begin{align}
\nu_k \triangleq \textrm{P}_{\textrm{no}, k} \bar{R}. \label{CapacityWithOutageDef}
\end{align}

For reliable AP-UE communication, we assume that a minimum NOP $\bar{\textrm{P}}_{\textrm{no}}$ is required for all UEs, i.e., $\textrm{P}_{\textrm{no}, k} \geq \bar{\textrm{P}}_{\textrm{no}}, \forall k \in \mathcal{K}$. 
If UE $k$ is served by AP only, then its NOP is given by
\begin{align}\label{PnonOutOfAPonlyDef}
\textrm{P}_{\textrm{no}, k} & = \mathbb{P}\Big\{ \frac{ p_k |h_{\textrm{d},k}|^2 }{ W } \geq \eta_0 \Big\} = \mathbb{P}\Big\{ \frac{ p_k g_{\textrm{d},k} \xi_{\textrm{d},k} }{ W } \geq \eta_0 \Big\} \notag \\
&= \mathbb{P}\Big\{ \xi_{\textrm{d},k} \geq \frac{ W \eta_0 }{ p_k g_{\textrm{d},k} } \Big\} \stackrel{(a)}{=} \exp\Big\{ - \frac{ W \eta_0 }{ p_k g_{\textrm{d},k} } \Big\},
\end{align}
where $(a)$ is due to $\xi_{\textrm{d},k}$ with exponential distribution.

On the other hand, if AP-UE $k$'s communication is also assisted by IRS $m_k$, its NOP can be obtained in closed-form by integrating over the probability density functions (pdf) of $|h_{\textrm{ir},k}|$ and $|h_{\textrm{d}, k}|$, whose exact expression is omitted here for brevity.
Although we can obtain its NOP in closed-form, it is still complicated when performing the inverse operation to obtain the required power $p_k$ that satisfies a certain NOP constraint.
To tackle this difficulty, we approximate the distribution of $Z_k^2$ by the Gamma distribution $\Gamma[\alpha, \beta]$ as in \cite{JiangbinLyuSpatialThroughput} with the \emph{shape} parameter $\alpha \triangleq (\mathbb{E}\{Z_k^2\})^2 / \textrm{var}\{Z_k^2\}$ and the \emph{inverse scale} parameter $\beta \triangleq \mathbb{E}\{Z_k^2\} / \textrm{var}\{Z_k^2\}$.
As a result, the NOP for UE $k$ served by IRS $m_k$ is then given by
\begin{small}
\begin{align}\label{PnonOutOfIRSaidedExplicit}
\textrm{P}_{\textrm{no}, k} \!=\! \mathbb{P}\big\{ Z_k^2 \geq \frac{W\eta_0}{p_k} \big\} \!\approx\! \frac{1}{\Gamma(\alpha)} \int_{\beta \frac{W \eta_0}{p_k}}^{\infty} t^{\alpha -1} e^{-t} \diff t \!=\! G_{\alpha}( \beta \frac{W \eta_0}{p_k} ),
\end{align}
\end{small}\noindent
where $\Gamma(\alpha) = \int_0^{\infty} t^{\alpha -1} e^{-t} \diff t$ is a constant, and $G_{\alpha}(\cdot)$ denotes the upper incomplete gamma function. 
As a result, for a given common rate $\bar{R}$ and the minimum NOP $\bar{\textrm{P}}_{\textrm{no}}$, by letting $\textrm{P}_{\textrm{no}, k} \geq \bar{\textrm{P}}_{\textrm{no}}$ in \eqref{PnonOutOfIRSaidedExplicit}, we can obtain the minimum required TP $p_k$ as
\begin{align}
p_k = W \eta_0 \beta/G_{\alpha, \textrm{inv}}(\bar{\textrm{P}}_{\textrm{no}}), \label{p_kIRSaided}
\end{align}
where $G_{\alpha, \textrm{inv}}(\cdot)$ denotes the inverse upper incomplete gamma function which is available in MATLAB.

Finally, it can be seen that the NOP in \eqref{PnonOutOfIRSaidedExplicit} is affected by the TP $p_k$ as well as the Gamma distribution parameters $\alpha$ and $\beta$, which in turn rely on the mean channel power gains of the IRS-related links.
Therefore, the IRS locations need to be jointly optimized along with the power allocation for the UEs.
%


\section{Problem Formulation and Proposed Solution}\label{ProblemFormulation}
We target at the system-level performance optimization for the single-cell multi-user system aided by multiple IRSs. 
Specifically, given the minimum NOP $\bar{\textrm{P}}_{\textrm{no}}$ requirement and the maximum TP constraint at AP, we aim to maximize the minimum throughput $\bar{\nu} \triangleq \bar{\textrm{P}}_{\textrm{no}} \bar{R}$ of all UEs by jointly optimizing the IRS locations, the UE-to-IRS association, and the power allocation among the UEs. 
The problem can be formulated as
\begin{align}
\mathrm{(P1)}:& \underset{
	\begin{subarray}{c}
	\bar{R},p_k, \bold{w}_m, a_{m, k}\\
	k \in \mathcal{K}, m \in \mathcal{M}
	\end{subarray}
}{\max} \quad \bar{\nu} \notag\\
\text{s.t.}\quad& \textrm{P}_{\textrm{no}, k} \geq \bar{\textrm{P}}_{\textrm{no}},\quad \forall k \in \mathcal{K}, \label{NonOutageProbabilityConstraintInP1} \\
& t_0 \textstyle{\sum}_{k=1}^{K} p_k \leq E_{\textrm{total}}, \label{TotalPowerConstraintInP1} \\
&a_{m, k} \in \{ 0, 1\}, m \in \mathcal{M}, k \in \mathcal{K}, \label{AssociationConstraint} \\
&K_m \leq n_{\textrm{t}}, \forall m \in \mathcal{M}, \label{ServedUEsNumberConstraintOfEachIRSinP1}
\end{align}
where $t_0 \triangleq T/n_{\textrm{t}}$ denotes the time slot duration, and $E_{\textrm{total}}$ denotes the AP's total transmitting energy budget in one time frame. 
Note that we choose the (average) throughput with the NOP requirement as the performance metric instead of the instantaneous rate per time slot, since we are interested in the system-level throughput optimization/capacity planning in the long run, which pertains to statistical CSI and UE distributions.
For the same reason, we consider the ``slow" power control policy where the TP $p_k$ for each individual UE $k$ is chosen based on the average channel statistics instead of fast adaptation to the fading state per time slot.
Likewise, we assume for simplicity that a feasible RB allocation solution is in place to arrange the group of UEs served by each IRS on the whole RB table.\footnote{Simple heuristics can be designed by ordering the UE groups based on $K_m$, which can then be sequentially arranged on the RB table. More advanced RB allocation design is left for future work.}

However, even considering only macro-decisions of IRS deployment and slow power control, the resulted problem (P1) is still difficult to solve due to the non-convex constraint \eqref{NonOutageProbabilityConstraintInP1} and the integer constraint \eqref{AssociationConstraint} with drastically increased complexity with the number of IRSs/UEs.
To tackle this challenge, we first investigate the impact of IRS deploying position on the AP's coverage range extension. Moreover, based on the derived NOP in \eqref{PnonOutOfIRSaidedExplicit} and the closed-form TP expression in \eqref{p_kIRSaided} for the IRS-served UE $k$, we can devise a power control policy that satisfies constraints \eqref{NonOutageProbabilityConstraintInP1} and \eqref{TotalPowerConstraintInP1}.
As a result, we are able to design a joint IRS deployment and AP power control scheme to obtain an efficient sub-optimal solution to (P1).


\subsection{Impact of the IRS Deploying Position in the Cell}\label{ImpactOfIRSdeployingPosition}
\begin{figure}[htbp]
	\centering
	\includegraphics[width=1\linewidth,  trim=5 0 30 0,clip]{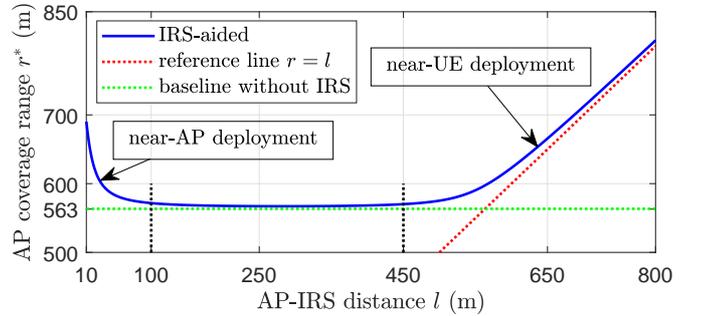}
	\caption{AP's coverage range $r^*$ under different AP-IRS distance $l$.\vspace{-2ex}}\label{ImpactOfIRSdeployingPositionNumericalResults}
\end{figure}

For the purpose of exposition, we consider a simplified setup where the UE is at horizontal distance $r$ from AP, and the IRS is deployed in between them with $l$ and $d$ denoting the AP-IRS and IRS-UE horizontal distances, respectively, and $r=l+d$.
Note that a similar setup is considered in \cite{QingqingWuTutorial}, which yet neglects the impact of the direct AP-UE path.
Consider a TP level of $p = 10$ dBm for the UE, and a coverage threshold in terms of the average received SNR at the UE, e.g., $\bar{\gamma} = 10$ dB.  Other parameters are given in Section \ref{NumericalResults}.
In the baseline case without IRS, the AP's coverage range (i.e., the maximum AP-UE distance $r^*$ that satisfies the coverage threshold $\bar{\gamma}$) is equal to $r^*=563$ m, which is indicated by a horizontal dotted line in Fig. \ref{ImpactOfIRSdeployingPositionNumericalResults}.
With the aid of the IRS, based on the average channel power gain $\mathbb{E}\{Z_k^2\}$ in \eqref{E_Z2}, the AP's coverage range $r^*$ under a given AP-IRS distance $l$ is also plotted in Fig. \ref{ImpactOfIRSdeployingPositionNumericalResults}.
Compared to the baseline case without IRS, it can be seen that the IRS helps extend the AP's coverage range $r^*$ mainly when the IRS is close to AP (i.e., small $l$ and large $d$) or close to the target UE (i.e., large $l$ and small $d$), e.g., when $l<100$ m or $l>450$ m, while the benefit brought by IRS is relatively small in the middle range, e.g., when $100<l< 450$ m.
The above observations suggest that the extended AP coverage range depends on the IRS position in the cell, and we find two favorable IRS deployment modes: 1) \textit{near-AP deployment} with long-range coverage, and 2) \textit{near-UE deployment} with local coverage.
This is fundamentally due to the average channel power term $g_{\textrm{i},k}g_{\textrm{r},k}$ of the AP-IRS-UE path in \eqref{E_Z2},
which follows the product-distance/double path-loss rule and becomes maximal when the IRS is placed near AP or the target UE.
%
%
%

\vspace{-2ex}

\subsection{Ring-Based IRS Placement Scheme}\label{RingBasedIRSplacementOptimization}
Motivated by the above insights, we propose a \textit{ring-based IRS placement scheme} where the UEs are grouped into ring regions based on the AP-UE distance, which are served by IRSs arranged on circles of different radius from AP, as shown in Fig. \ref{IRS}.
Since the cell-edge UEs typically present as the performance bottleneck, they are preferentially served by IRSs in order to maximize the minimum throughput in the cell.
Specifically, the $M$ IRSs are divided into $I$ disjoint subsets, denoted by $\mathcal{M}_i$, $i\in\mathcal{I} \triangleq \{1, 2, \cdots, I\}$, each with $M_i\triangleq |\mathcal{M}_i|$ IRSs which are arranged on a circle of radius $L_i$ centered at AP and are responsible for serving the UEs in the ring region $\mathcal{S}_i$ within distance range $(R_{\textrm{in}, i}, R_{\textrm{in}, i-1}]$ from AP.
Note that the UEs closer to the cell edge are served first, by the IRS subset with a smaller index, e.g., the ring regions $\mathcal{S}_1$, $\mathcal{S}_2$ and $\mathcal{S}_3$ are served by the IRS subsets with radius $L_1$, $L_2$ and $L_3$ in Fig. \ref{IRS}, respectively.
By default, the cell-edge UEs are served by IRS subset $\mathcal{M}_1$, i.e., $R_{\textrm{in}, 0}=R_{\textrm{ex}}$.\footnote{In general, for the whole IRS-served region, the outer range $R_{\textrm{in}, 0}\leq R_{\textrm{ex}}$ and can also be optimized, which is investigated in Section \ref{NumericalResults}.}
For the rest of UEs in the inner disc region $\mathcal{S}_0$ within distance range $[0, R_{\textrm{in}, I}]$, they are served by AP only.


For each ring region $\mathcal{S}_i$, the serving IRSs can be deployed using either the near-AP mode or near-UE mode.
However, considering the limited space and site availability near AP, for the purpose of illustration, we consider only one IRS subset (e.g., $\mathcal{M}_1$) using the near-AP mode in this paper, whereby at most $M_{1,\textrm{max}}$ IRSs can be placed at $L_\textrm{min}$ m away from AP with equal angular spacing, i.e., $L_1=L_\textrm{min}$ and $M_1 \leq M_{1,\textrm{max}}$.
For other IRS subsets $\mathcal{M}_i$, $i\in\mathcal{I}\setminus \{1\}$, the near-UE mode is adopted, where the $M_i$ IRSs are placed with equal angular spacing and radius $L_i$, which is set as $L_i=( R_{\textrm{in}, i} + R_{\textrm{in}, i-1} ) / 2$ for simplicity.
As a result, the region served by each IRS in $\mathcal{M}_i$ is an \textit{annulus sector}\footnote{For the case with non-uniform UE distribution, the shape and size of each annulus sector can be adjusted to fit the local UE density. Moreover, the practical IRS locations can be adjusted based on their nearby mounting infrastructure.} from the radius $R_{\textrm{in}, i}$ to $R_{\textrm{in}, i-1}$ with a central angle $\phi_i \triangleq 2\pi/M_i$ as shown in Fig. \ref{IRS}, whose area is given by $A_i\triangleq  \pi ( R_{\textrm{in}, i-1}^2 - R_{\textrm{in}, i}^2 )/M_i$ with an average number of UEs given by $\bar{K}_{i} =  \lambda A_{i}$.

Note that the number of UEs supported by each IRS is usually finite due to practical factors such as the limited number of time slots $n_\textrm{t}$ per time frame.
On the other hand, the number of UEs in each IRS-served region may vary depending on the random realization of UE locations.
Therefore, in order to satisfy constraint \eqref{ServedUEsNumberConstraintOfEachIRSinP1} most of the time, we place a limit $\bar{K}_{\textrm{IRS}}$ on the \textit{average} number of UEs served by each IRS, which is sufficiently smaller than $n_t$, i.e., $\bar{K}_{i}\leq \bar{K}_{\textrm{IRS}} \ll n_{\textrm{t}}$.
In the unlikely case that the number of UEs within each IRS-served region still exceeds $n_{\textrm{t}}$, the $n_{\textrm{t}}$ UEs closer to the IRS are served preferentially while the rest of UEs are served by AP only.
Similarly, in order to satisfy the total energy constraint \eqref{TotalPowerConstraintInP1} statistically, we first allocate a power ratio of $\rho_i>0$ to the UE region $\mathcal{S}_i$, $i=\mathcal{I}\cup \{0\}$, with $\sum_{i=0}^I \rho_i =1$.
As a result, the total energy allocated to the UE region $\mathcal{S}_i$ is given by $E_i\triangleq \rho_i E_{\textrm{total}}$, which is then allocated to the UEs in that region based on a certain power control policy (discussed later in Section \ref{SectionPowerControl}), with the objective to equalize the UEs' average NOP, denoted by $\bar{\textrm{P}}_{\textrm{no},i}$.
As a result, in our proposed ring-based IRS placement framework, the original problem (P1) can be reduced to
\begin{align}
\mathrm{(P2)}:&\underset{
	\begin{subarray}{c}
	\bar{R}, \rho_0,\rho_i, M_i, R_{\textrm{in}, i},i \in \mathcal{I}
	\end{subarray}
}{\max} \quad\bar{\nu} \notag\\
\text{s.t.}\quad&\bar{\textrm{P}}_{\textrm{no},i} \geq \bar{\textrm{P}}_{\textrm{no}}, \forall i \in \mathcal{I}\cup \{0\},\label{NonOutageProbabilityConstraintInP2}\\
& \textstyle{\sum}_{i=0}^I \rho_i =1, \quad (\rho_i\geq 0, \forall i \in \mathcal{I}\cup \{0\}) \label{TotalPowerConstraintInP2} \\
&\textstyle{\sum}_{i=1}^{I} M_i = M, (M_i \in \mathbb{N}, M_1 \leq M_{1,\textrm{max}}), \label{TotalIRSsNumberConstraint}\\
&\bar{K}_{i} \leq \bar{K}_{\textrm{IRS}}, \forall i \in \mathcal{I}, \label{AverageServedUEsNumberConstraintOfEachIRSinP2}\\
&0 \leq R_{\textrm{in}, I} \leq \cdots \leq R_{\textrm{in}, 2} \leq R_{\textrm{in}, 1} \leq R_{\textrm{ex}}, \label{RinConstraint}
\end{align}
where $\mathbb{N}$ denotes the set of natural numbers.

Compared to (P1), the problem size has been reduced from $O(K\times M)$ to $O(I)$, whereby the UE-to-IRS association is now specified by the IRS-served annulus sectors, and the power control for all UEs is reduced to the problem of equalizing the average NOP in each ring region.
Therefore, for the case with a small number of IRSs $M$ and/or a small number of rings $I$, an optimal solution to (P2) can be found by linear search over $R_{\textrm{in}, i}$, $M_i$ and $\rho_i$, $i \in \mathcal{I}$.
However, as the number of rings $I$ increases, the searching process becomes increasingly inefficient.
To this end, an efficient searching algorithm is proposed in Algorithm \ref{AlgEfficientSearching} to obtain a close-to-optimal solution to (P2) for arbitrary $I$. This is motivated by the solutions of optimal line search in Section \ref{NumericalResults}, which favors the near-AP mode in terms of coverage performance, and also suggests that the resulted annulus sectors served by near-UE IRSs tend to have similar shape and size (i.e., with a similar radial/arc dimension) regardless of the ring they reside.
Therefore, in Algorithm \ref{AlgEfficientSearching}, the IRSs fill up the near-AP positions first (Steps 1 to 4), while the rest of IRSs are deployed near UEs (Steps 5 to 16). 
For the near-UE IRSs, they are first deployed to cover more UEs using the maximum allowed $\bar{K}_{\textrm{IRS}}$ as in \eqref{AverageServedUEsNumberConstraintOfEachIRSinP2}, and then get denser to get closer to their served UEs if more IRSs are available (Step 6). 
Moreover, the region served by near-UE IRSs are divided into $I-1$ equal-interval rings and then filled up by the annulus sectors of the same area $A_i\!\gets\!\bar{K}_i/\lambda$ (Steps 7 to 11).
As a result, under given IRS deployment, power control can be performed as in Section \ref{SectionPowerControl} to obtain the power allocation ratios $\rho_i, i=\mathcal{I}\cup \{0\}$ to maximize the common throughput $\bar{\nu}$ (Steps 2 and 12).
Finally, by searching over a maximum of $I_\textrm{max}$ rings and recording the highest $\bar{\nu}$ and its associated solution, we can find an efficient close-to-optimal solution to (P2) with further reduced complexity.\footnote{There is no need of line searching for $R_{\textrm{in}, i}$ and $M_i$, $i \in \mathcal{I}$.}
Next, we introduce the power control policy used in the above algorithms.



\begin{algorithm}[H]\caption{Efficient Searching Algorithm for Achieving Max-min Throughput in (P2)}\label{AlgEfficientSearching}
	\begin{small}
		\textbf{Input:} Number of IRSs, $M$; maximum number of rings, $I_{\textrm{max}}$.\\
		\textbf{Output:} Common throughput $\bar{\nu}^*$ achieved by solution $\rho_i^*$, $M_i^*$, $R_{\textrm{in}, i}^*$, $1 \leq i \leq I^*$, with $I^*$ being the number of rings adopted.
		\begin{algorithmic}[1]
			\IF{$M \leq M_{1,\textrm{max}}$}
			\STATE Deploy all IRSs near AP, i.e., $I^*\!\gets\!1$, $M_1^*\!\gets\!M$. Let $\bar{K}_1\!\gets\!\bar{K}_{\textrm{IRS}}$. Obtain $A_1\!\gets\!\bar{K}_1/\lambda$ and hence $R_{\textrm{in}, 1}^*$. Obtain $\rho_1^*$ by power control in Sec. \ref{SectionPowerControl} to maximize common throughput $\bar{\nu}^*$.
			\ELSE
			\STATE Initialize $\bar{\nu}^*\!\gets\!0$. Deploy $M_1 \!=\! M_{1,\textrm{max}}$ IRSs near AP. Let $\bar{K}_1\!\gets\!\bar{K}_{\textrm{IRS}}$ and obtain $R_{\textrm{in}, 1}$. Deploy the rest of IRSs near UEs.
			\FOR{$I = 2,\cdots, I_{\textrm{max}}$}
			\STATE Let $\bar{K}_i\!=\!\bar{K}_{\textrm{IRS}}, i\!=\!2,\cdots,I$. Obtain $R_{\textrm{in}, I}$ based on the total serving area $A\!=\!\sum_{i\in\mathcal{I}} M_i A_i$ of all IRSs. If $A\!>\!\pi R_\textrm{ex}^2$, let $R_{\textrm{in}, I}\!\gets\!0$ and update $\bar{K}_i\!\gets\!\frac{\lambda \pi (R_{\textrm{in}, 1}^2 \!-\! R_{\textrm{in}, I}^2)}{M-M_1}, i\!=\!2,\cdots,I$.
			\FOR{$i = 2,\cdots, I$}
			\STATE Divide $[R_{\textrm{in}, I}, R_{\textrm{in}, i-1}]$ into $I \!-\! i \!+\! 1$ equal intervals with width $\delta\triangleq \frac{(R_{\textrm{in}, i-1}-R_{\textrm{in}, I})}{ (I-i+1) }$. Let $R_{\textrm{in}, i} \gets R_{\textrm{in}, i-1} \!-\! \delta$. 
			\STATE Let $M_i\!\gets\!\lceil \lambda \pi (R_{\textrm{in}, i-1}^2 \!-\! R_{\textrm{in}, i}^2)/\bar{K}_i \rceil$, and $A_i\!\gets\!\bar{K}_i/\lambda$.
			\STATE Refine $R_{\textrm{in}, i}$ based on the area $M_i A_i$ of ring region $\mathcal{S}_i$.
			\ENDFOR
			\STATE Power control in Sec. \ref{SectionPowerControl} to obtain common throughput $\bar{\nu}$.
			\IF{$\bar{\nu} > \bar{\nu}^*$}
			\STATE $\bar{\nu}^* \gets \bar{\nu}$, $I^* \gets I$, and update $\rho_i^*$, $M_i^*$, $R_{\textrm{in}, i}^*$, $1 \leq i \leq I^*$.
			\ENDIF
			\ENDFOR
			\ENDIF
		\end{algorithmic}
	\end{small}
\end{algorithm}

\subsection{Power Control Policy}\label{SectionPowerControl}
The power control subproblem for each UE region $\mathcal{S}_i$, $i=\mathcal{I}\cup \{0\}$, aims to equalize the UEs' NOP in that region such that its average NOP $\bar{\textrm{P}}_{\textrm{no},i}$ satisfies constraint \eqref{NonOutageProbabilityConstraintInP2} and an average common (minimum) throughput $\bar{\nu} \triangleq \bar R_i \bar{\textrm{P}}_{\textrm{no}}$ in that region can be found, subject to the total energy budget $E_i\triangleq \rho_i E_{\textrm{total}}$.
Based on the NOP expressions derived in Section \ref{NonOutageProbability}, we propose power control policies to achieve the above objective for the case with and without IRS, respectively.
 
\subsubsection{AP-Only Region $\mathcal{S}_0$}\label{APonlyRegionPowerAllocationMethod}
Based on the NOP in \eqref{PnonOutOfAPonlyDef} for the AP-only case, in order to equalize the NOP in the region $\mathcal{S}_0$, we assume that AP adopts the ‘‘slow’’ channel inversion power control (CIPC) based on the average channel power gain $g_{\textrm{d},k}$ such that the received SNR of each UE $k$ is equal to a common value $\bar{\gamma}$, i.e., $\gamma_k = p_k g_{\textrm{d},k} / W=\bar{\gamma}$. As a result, for UE $k$ at distance $r_k$ from AP, its allocated TP is given by
\begin{equation}\label{PowerAP}
	p(r_k)=\frac{\bar{\gamma} W}{g_{\textrm{d},k}}=\frac{\bar{\gamma} W}{\alpha_0 (r_k^2+H_\textrm{A}^2)^{-n_0/2}},
\end{equation}
Based on such a power control policy, the average consumed energy within one time frame is given by
\begin{small}
\begin{equation}
\bar E_0\triangleq \lambda \int_{\varphi = 0}^{2\pi}\int_{r = 0}^{R_{\textrm{in}, I}} p(r) t_0 r \diff r \diff\varphi, \label{TotalPowerConstraintOfAPonlyChannelInversion}
\end{equation}
\end{small}%
where $(r, \varphi)$ denotes the polar coordinate of UE $k$ centered at AP.
By substituting \eqref{PowerAP} into \eqref{TotalPowerConstraintOfAPonlyChannelInversion} and letting $\bar E_0=E_0=\rho_0 E_\textrm{total}$, 
we can obtain the average common SNR as
\begin{align}
\bar{\gamma} = \frac{\alpha_0 \rho_0 E_\textrm{total}}{ 2 \pi \lambda W t_0 F_0(R_{\textrm{in}, I}) }, \label{averageSNROfAPonlyRegion}
\end{align}
with $F_0(R_{\textrm{in}, I}) \triangleq \int_0^{R_{\textrm{in}, I}} ( r^2 + H^2_{\textrm{A}} )^{n_0/2}r \diff r$.

Based on \eqref{PnonOutOfAPonlyDef} and \eqref{PowerAP}, the NOP of UE $k$ in the AP-only region is given by
\begin{align}
\textrm{P}_{\textrm{no}, k} = \exp\bigg\{ - \frac{ W \eta_0 }{ \frac{\bar{\gamma} W}{ g_{\textrm{d},k}} g_{\textrm{d},k} } \bigg\} =\exp\Big\{ \frac{-\eta_0}{\bar{\gamma}} \Big\}\triangleq \bar{\textrm{P}}_{\textrm{no},0},\label{PnonOutOfAPonlyExplicit}
\end{align}
where $\bar{\textrm{P}}_{\textrm{no},0}$ denotes the common NOP in the AP-only region $\mathcal{S}_0$, and $\eta_0 = 2^{ \bar{R}_0 }-1$ is the SNR threshold for the outage event.
Therefore, by letting $\bar{\textrm{P}}_{\textrm{no},0}=\bar{\textrm{P}}_{\textrm{no}}$, we can obtain the maximum common rate $\bar{R}_0$ and hence the maximum common throughput in the AP-only region as
\begin{align}
\bar{\nu}_0 \triangleq \bar{\textrm{P}}_{\textrm{no}} \bar{R}_0 = \bar{\textrm{P}}_{\textrm{no}} \log_2( 1+\bar{\gamma}\ln( 1/\bar{\textrm{P}}_{\textrm{no}} ) ). \label{maxRbarOfAPonlyRegion}
\end{align}

\subsubsection{IRS-Aided Region $\mathcal{S}_i$}\label{IRSaidedRegionPowerAllocationMethod}
Based on the Gamma approximation for the IRS-assisted signal power gain $Z_k^2$ in Section \ref{NonOutageProbability}, we have derived a closed-form expression in \eqref{p_kIRSaided} for the required TP $p_k$ in achieving the minimum NOP $\bar{\textrm{P}}_{\textrm{no}}$, under given locations of UE $k$ and its serving IRS.
Based on such a power control policy, the average consumed energy within one time frame in the IRS-aided region $\mathcal{S}_i$ is given by
\begin{small}
\begin{align}
\bar E_i\triangleq M_i \lambda \int_{\varphi = 0}^{\phi_i}\int_{ r = R_{\textrm{in}, i} }^{ R_{\textrm{in}, i-1} } p_k t_0 r \diff r \diff\varphi = M_i \lambda W \eta_0 t_0 F_i, \label{TotalPowerConstraintOfIRSaidedRegion}
\end{align}
\end{small}%
where $F_i \triangleq \int_{\varphi = 0}^{\phi_i}\int_{ r = R_{\textrm{in}, i} }^{ R_{\textrm{in}, i-1} } \frac{ \beta }{ G_{\alpha, \textrm{inv}}( \bar{\textrm{P}}_{\textrm{no}} ) } r \diff r \diff\varphi$ and the Gamma distribution parameters $\alpha$ and $\beta$ depend on the UE coordinate $(r,\varphi)$ via \eqref{E_Z2} and \eqref{var_Z2}.
Based on the SNR threshold $\eta_0 = 2^{ \bar{R}_i }-1$,
by letting $\bar E_i=E_i=\rho_i E_\textrm{total}$, we can obtain the maximum common throughput in the IRS-aided region $\mathcal{S}_i$ as
\begin{align}
\bar{\nu}_i \triangleq \bar{\textrm{P}}_{\textrm{no}} \bar{R}_i = \bar{\textrm{P}}_{\textrm{no}} \log_2\Big( 1+\frac{ E_{i} }{ M_i \lambda W t_0 F_{i} } \Big). \label{maxRbarOfIRSaidedRegionS2i}
\end{align}

Finally, by equalizing the common throughput of all UE regions $\mathcal{S}_i$, $i \in \mathcal{I}\cup \{0\}$, their optimal power allocation ratios can be obtained in closed-form, which is omitted here for brevity.

\section{Numerical Results}\label{NumericalResults}

This section verifies our analytical results for the achieved common throughput $\bar{\nu}^*$ by Monte Carlo (MC) simulations.
Each MC result is obtained by averaging over 1000 randomly generated topologies, with $10^6$ fading realizations per channel.
The following parameters are used: $R_{\textrm{ex}}=250$ m,
$H_\textrm{A} = 10$ m, $H_\textrm{I} = 1$ m, $E_\textrm{total} = 10^{-3}$ J,  $N_0 = -174$ dBm/Hz, $f_c = 2$ GHz, $n_0 = 3$, $K = 500$, $\bar{\textrm{P}}_{\textrm{no}} = 0.95$, $B = 5$ MHz, $t_0 = 0.5$ ms, $n_{\textrm{t}} = 20$, $n_{\textrm{b}} = 25$, $N=2000$, $L_\textrm{min}=10$ m, $M_{1,\textrm{max}}=10$ and $\bar{K}_{\textrm{IRS}} = 10$.

\begin{figure}
	\centering
	\includegraphics[width=1\linewidth,  trim=10 0 10 0,clip]{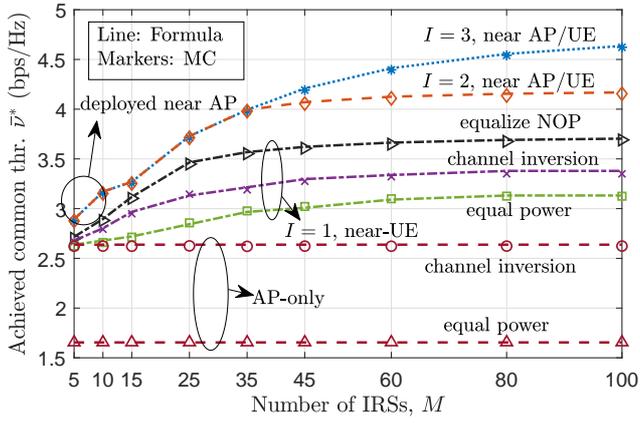}
	\caption{Achieved common throughput under different schemes.\vspace{-2ex}}\label{MaxMinThroughput}
\end{figure}

For small/moderate number of IRSs/rings (e.g., $M\!\leq\! 100$, $I\!=\!1, 2, 3$), the max-min throughput of our proposed scheme can be obtained by optimal line search as discussed in Section \ref{RingBasedIRSplacementOptimization}, which is plotted in Fig. \ref{MaxMinThroughput},
along with the results obtained by the benchmark scheme without IRS and/or with other power control policies. 
First, it can be seen that the analytical results match well with the MC results.
Second, our proposed scheme significantly outperforms the baseline scheme without IRS, achieving $180.19\%$ and $75.77\%$ higher common throughput than the AP-only scheme with equal power allocation or slow CIPC, respectively, with $M=100$ IRSs and $I=3$ rings.
Moreover, our proposed power control policy that equalizes the NOP for the IRS-aided region further improves over the two benchmark power control policies.
In addition, for our proposed scheme with hybrid near-AP and near-UE deployment options (e.g., $I=2$ or $I=3$), it is found that the near-AP positions are filled up first before the rest of IRSs are deployed in the near-UE mode.
This could be attributed to the wide coverage range of the near-AP IRS that could serve distant UEs (which typically suffer from severe path-loss) spreading in a wide angular region, as shown by the red shaded region in Fig. \ref{IRS}.
On the other hand, by optimal line search, it is found that the resulted annulus sectors served by near-UE IRSs tend to have similar shape and size (i.e., with a similar radial/arc dimension) regardless of the ring they reside.
The underlying reason is two-fold: 1) the size of each annulus sector is constrained by the maximum number of supported UEs per IRS; 2) the shape of the UE region served by a near-UE IRS tends to have a minimal dimension in order to minimize the worst-case distance to its served UEs.
These observations also motivate our design of Algorithm \ref{AlgEfficientSearching}.

In addition, for our considered setup with different cell radius $R_{\textrm{ex}}$, we have also searched over the exterior range $R_{\textrm{in},0}$ of the whole IRS-aided region, and found that the optimal solution is always to serve the cell-edge UEs first, i.e., $R_{\textrm{in},0}=R_{\textrm{ex}}$, regardless of the number of rings deployed.
This could be due to the fact that the cell-edge UEs are the most power-limited and would thus help improve the max-min throughput when they are preferentially served by IRSs.

\begin{figure}
	\centering
	\includegraphics[width=1\linewidth,  trim=10 0 10 0,clip]{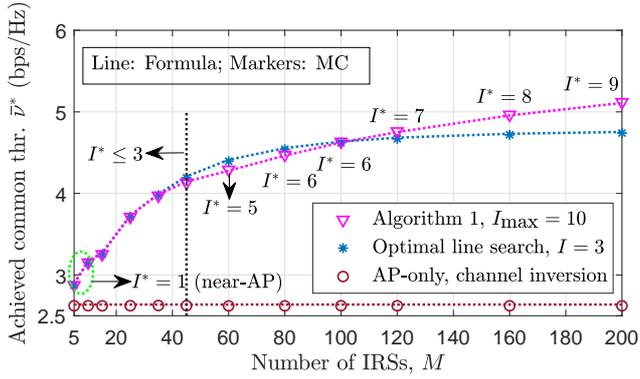}
	\caption{Comparison of the optimal line search and Algorithm 1.\vspace{-2ex}}\label{HeuristicVSExhaust}
\end{figure}

Finally, as the number of IRSs further increases, the optimal line search with more rings becomes inefficient, whereby our proposed Algorithm 1 helps in this case to obtain a close-to-optimal solution efficiently, as shown in Fig. \ref{HeuristicVSExhaust}.
It can be seen that Algorithm 1 achieves near-optimal performance when the number of IRSs is small (e.g., $M \leq 45$).
In addition, as the number of IRSs increases beyond 100, Algorithm \ref{AlgEfficientSearching} outperforms the optimal line search with $I=3$ by allowing searching for more rings (e.g., $I_\textrm{max}=10$).
Moreover, it is observed that the IRSs first tend to spread out over the cell to cover more UEs, and then get denser to get closer to their served UEs and achieve higher throughput, as the number of IRSs increases.

\section{Conclusions}
This paper investigates the achievable max-min throughput of a single-cell multi-IRS-assisted multiuser system by joint IRS placement and AP power control, which is shown to be a mixed-integer non-convex problem with drastically increased complexity as the number of IRSs/UEs increases.
To tackle this challenge, we first derive the IRS-aided channel power statistics and obtain a closed-form approximation of the required TP to achieve a certain NOP under given UE and IRS locations.
In addition, two desirable modes of IRS deployment are observed, i.e., near-AP deployment with long range coverage, and near-UE deployment with local coverage.
Thereby, a ring-based IRS placement scheme with reduced complexity is proposed along with a power control policy that equalizes the UEs' NOP for achieving the average max-min throughput.
An efficient searching algorithm is further proposed to obtain a close-to-optimal solution for arbitrary number of IRSs/rings.
Numerical results validate our analysis and show that our proposed scheme significantly outperforms the benchmark schemes without IRS and/or with other power control policies.

\vspace{-1ex}

\bibliography{IEEEabrv,BibIRSdeployment}
\end{document}